# Continuous entanglement distribution over a transnational 248 km fibre link

Sebastian Philipp Neumann[1,2] *, Alexander Buchner[1,2], Lukas Bulla[1,2], Martin Bohmann[1,2] & Rupert Ursin[1,2] *

[1] *Institute for Quantum Optics and Quantum Information Vienna, Boltzmanngasse 3, 1090 Vienna*
[2] *Vienna Center for Quantum Science and Technology, Boltzmanngasse 5, 1090 Vienna, Austria*

\* Corresponding authors: sebastian.neumann@oeaw.ac.at, rupert.ursin@oeaw.ac.at

**Entanglement is the basis of many quantum applications [1–8]. The technically most mature of them, quantum key distribution [9–11], harnesses quantum correlations of entangled photons to produce cryptographic keys of provably unbreakable security [12, 13]. A key challenge in this context is the establishment of continuously working, reliable long-distance distributions of entanglement. However, connections via satellites [14, 15] don't allow for interruption-free operation, and deployed fibre implementations have so far been limited to less than 100 km by losses [16], a few hours of duty time [17, 18], or use trusted nodes [19]. Here, we present a continuously working international link between Austria and Slovakia, directly distributing polarization-entangled photon pairs via 248 km of deployed telecommunication fibre. Despite 79 dB loss, we measure stable pair rates of 9 s$^{-1}$ over an exemplary operation time of 110 hours. We mitigate multi-pair detections with strict temporal filtering, enabled by nonlocal compensation of chromatic dispersion. Fully automated active polarization stabilization keeps the entangled state's visibility at 86% for altogether 82 hours, producing 403 kbit of quantum-secure key at a rate of 1.4 bits/s. Our work paves the way for low-maintenance, ultra-stable quantum communication over long distances, independent of cloud coverage and time of day, thus constituting an important step towards the quantum internet.**

Fibre-based QKD systems offer stable operation, independence from meteorological conditions, substantially reduced maintenance effort and the use of already deployed telecommunication infrastructure. These advantages can compensate for their higher losses [20] compared to satellite connections. Therefore, while intercontinental quantum connections will most likely be operated using satellites, shorter distances of several hundred kilometres can be covered by fibre links [19, 21]. Metropolitan fibre networks deploying entanglement-based QKD additionally have the advantage of allowing to fully connect many users in a straightforward fashion [17, 22], potentially on fibres used for internet traffic, wavelength-multiplexed with the classical signal [23]. Nevertheless, losses in the fibres, imperfect preparation of entangled states, chromatic dispersion, polarization mode dispersion, and timing precision in the detection of single photons hinder stable operation over long distances.

Up until today, the longest distance for entanglement distribution in deployed fibre was along a single 96 km fibre between Malta and Sicily [16]; additionally, the same submarine cable was used for a round-trip connection of altogether 192 km [24]. The longest uninterrupted operation of entanglement distribution, using active stabilization, has been demonstrated to work for 6 hours along a deployed 10-km-link [18].

In this work, we combine state-of-the-art equipment and optimal exploitation of our polarization-entangled photons' quantum properties to demonstrate continuously operated entanglement distribution along a record distance of 248 km of deployed telecom fibre, connecting Bratislava in Slovakia and St. Pölten via Vienna in Austria. Additionally, we show, for the first time, ground-based entanglement distribution for QKD in a real-life two-channel configuration, while one channel crosses the international border between Austria and Slovakia without any intermediary trusted nodes. Despite unprecedented total loss of 79 dB, we achieve entangled pair rates of

9 s⁻¹ and secure key rates of 1.4 bits/s on average. We continuously operate the link for a record of 110 hours by actively stabilizing the polarization in a highly efficient, nonlocal way, achieving a duty cycle of 75% and a total key of 403 kbit.

**Sender and receiver infrastructure**

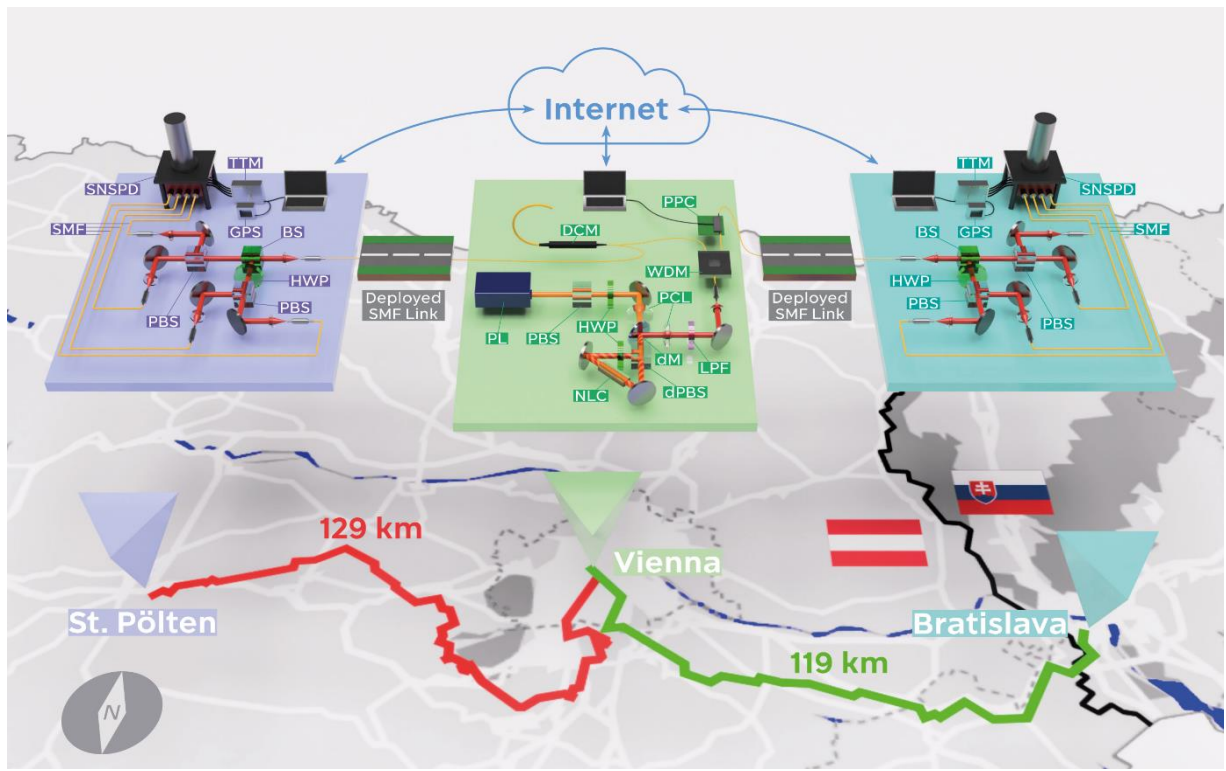

*Figure 1. Sketch of the Setup. The source of entangled photon pairs is situated in Vienna. We create polarization-entangled photon pairs at two distinct telecommunication wavelengths by pumping a non-linear crystal (NLC) in a Sagnac configuration with a 775 nm laser (PL) and collecting the down-converted photons with single-mode fibres (SMF) connected to a wavelength division (de-)multiplexer (WDM). The idler photon passes a dispersion compensation module (DCM) which nonlocally recovers the entangled state's tight temporal correlations broadened by chromatic dispersion along the link. The idler is then directed along 129 km of fibre to a polarization measurement module (PMM) in St. Pölten in Lower Austria. The signal photon passes an automatized in-fibre piezo-based polarization controller (PPC) which nonlocally realigns the phase of the entangled state, should its quality decrease. Afterwards, it travels to a PMM in Bratislava of the same design as the one in Austria. In these PMMs, the photons are randomly directed to orthogonal measurements in two mutually unbiased linear polarization bases. They photons impinge on superconducting nanowire single-photon detectors (SNSPD), and a GPS-clock-disciplined time-tagging module (TTM) records detection time, measurement basis and outcome. Via classical internet connections, the two measurement stations' detection events are compared and coincidences calculated. If their quantum bit error rate increases, Vienna starts the polarization alignment. PBS: polarizing beamsplitter, HWP: half-wave plate, PL: planoconvex lens, dPBS: dichroic PBS, DM: dichroic mirror, LPF: longpass filter, BS: 50:50 beamsplitter*

We implement a symmetric two-channel QKD system in a "source in the middle" configuration, following the BBM92 protocol [10] for polarization entanglement (see Fig. 1 and the Methods section). The source of entangled photon pairs is situated in Vienna, utilizing continuous-wave-pumped spontaneous parametric down-conversion

(SPDC) in a Sagnac configuration [25] and wavelength division demultiplexing (WDM) to create entangled photons of 100 GHz spectral width around 1550.12 nm with > 99% fidelity to the Bell state

$$|\phi^+\rangle = 1/\sqrt{2}\,(\,|H\rangle_{SP}|H\rangle_B + |V\rangle_{SP}|V\rangle_B) \qquad (1)$$

$$= 1/\sqrt{2}\,(\,|D\rangle_{SP}|D\rangle_B + |A\rangle_{SP}|A\rangle_B\,),$$

where H (V, D, A) refers to horizontal (vertical, diagonal, antidiagonal) polarization. The subscripts denote the receiver stations of the respective photon: St. Pölten in Lower Austria (SP) and the campus of the Slovakian Academy of Sciences in Bratislava (B). The connections are realized via deployed telecom fibres of 129 and 119 km length, respectively. The receivers measure each photon's polarization state using a bulk polarization measurement module in two mutually unbiased, randomly chosen linear polarization bases (H/V or, with equal probability, D/A). Superconducting nanowire single-photon detectors (SNSPD) connected to a time-tagging module (TTM) register each detection event. By comparing the detection times, SP and B identify the entangled photon pairs. The total transmission of each link was determined to be −40.2 dB (−38.4 dB) for the link to SP (B), including all losses and detection efficiencies. From this, it follows that an average of 8.9 pairs were detected per second between the receiver stations. We can only harness the quantum correlations of those pairwise (or "coincident") events measured in the same polarization basis. The rate of these photon pairs is called "sifted" key rate and amounted to 4.4 s$^{-1}$ on average in our case due to our passively implemented, balanced and random basis choice. Strict temporal filtering with a width of $t_{CC}$=114 ps, also called the "coincidence window", further reduces this value to 3.8 s$^{-1}$. Of this rate, on average 0.46 coincidences originate from accidental counts, contributing 4.4 percent points to the quantum bit error rate (QBER), which is well below the 11.0% limit necessary to arrive at a non-zero key [26].

**Nonlocal dispersion compensation**

To reach such a low level of accidental coincidences over our high-loss link, sufficiently high temporal detection precision is necessary, allowing for strict temporal filtering [27]. The greatest effect detrimental to timing precision in our experiment is chromatic dispersion (CD) along the fibres. CD causes photons with finite spectral distribution to disperse in time, thus spreading the entangled photons' temporal intensity correlation function $g^{(2)}$. In our case, optical time-domain reflectometer measurements of the link yielded a CD of 16.8 (6.0) ps/nm/km at 1550 nm for the link to St. Pölten (Bratislava). Thus, our 100-GHz-bandwidth photons would suffer from a total CD of about 1.8 ns over the full fibre stretch, which is 50 times larger than the SNSPD and TTM jitters combined. To prohibit this, we deployed a single passive dispersion compensation module (DCM) with equal and opposite CD of −1.8 ns acting on the photon traveling to St. Pölten only. Such nonlocal dispersion compensation [28–30] harnesses the intrinsic quantum properties of entangled photon pairs to narrow their $g^{(2)}$ distribution by acting on one photon of a pair only. The residual CD-induced temporal spread after compensation was masked by SNSPD jitter, TTM jitter and GPS clock drift, which we identify as the remaining contributions to the overall timing uncertainty (see Methods section).

**Active polarization stabilization**

Besides accidental coincidence counts, erroneous polarization measurements of (correctly identified) photon pairs contribute to the QBER. While such errors can be kept below 0.4% under laboratory conditions and for short time scales [31], the nature of our experiment required active polarization stabilization of the altogether 248 km of deployed fibres, which were subject to polarization drift [32]. Without stabilization, this drift randomly changes the phase of the Bell state in Eq. (1), such that the quantum correlations between the photons at SP and B can no longer be observed with sufficient fidelity. To guarantee long-term operation of our link despite this effect, we implemented an automatized algorithm for a piezo-based polarization controller (PPC) working on the fibre channel to Bratislava. It switched on whenever the QBER increased above 9% (see Fig. 2) and used the QBER value (calculated by B) directly as input, thus requiring no reference laser. Due to the nonlocal nature of the entangled quantum state, manipulation of just the photon in the B mode allows to arrive at the correlations of Eq. (1).

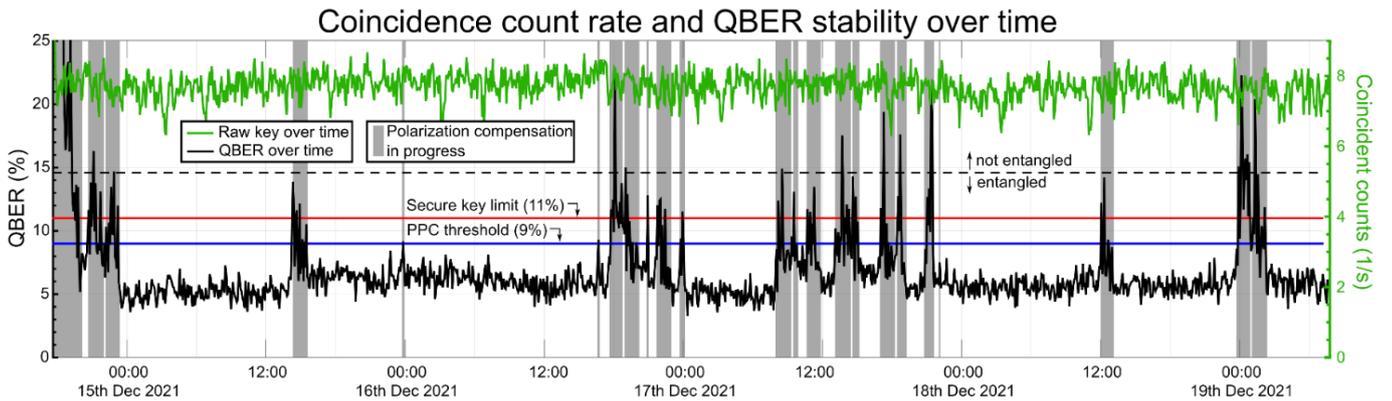

*Figure 2. Quantum bit error rate (QBER) and coincident counts over time with a coincidence window of tCC = 114 ps. When polarization drifts along the overland fibre link increase the QBER above 9%, the piezo-based polarization controller (PPC) starts an iterative alignment procedure to reduce the QBER below 7%. The limits were chosen such that alignment is started well before the 11 % limit of no key. Alignment takes between 8:10 minutes and 1:22 hours, and 57 minutes on average. We assume that the longer alignment phases are caused by polarization drift during alignment. The longest uninterrupted stable operation time amounted to 16:36 hours. The coincidence counts stay at a constant value of around 7.7 s−1 over the full 110 hours for tCC = 114 ps.*

The PPC iteratively scanned the voltage for each of the four fibre-squeezing piezoelectric crystals, thus optimizing the QBER via a hill-climb algorithm. Due to the low coincidence rates, it took the algorithm 2:32 minutes on average to determine the QBER value with a precision of ±0.2%. Therefore, the mean length of the alignment procedure amounted to 57 minutes along the link, rather than sub-seconds in the laboratory, where coincidence rates were in the order of $10^4$ s$^{−1}$. While the PPC is in operation, no key can be created, since both bit and basis information are sent via a classical internet connection. In our case, the PPC was active for altogether 27:57 of 109:55 hours, i.e., 25.4% of the time. This is equivalent to a duty cycle of 74.6% for the whole QKD scheme, with the longest uninterrupted operation time being 16 hours 37 minutes. During this duty cycle, the QBER was kept at an average of 7.0%, where we attribute 2.6% to polarization measurement errors and 4.4% to accidental coincidences (see Methods section).

**Secure key rate analysis**

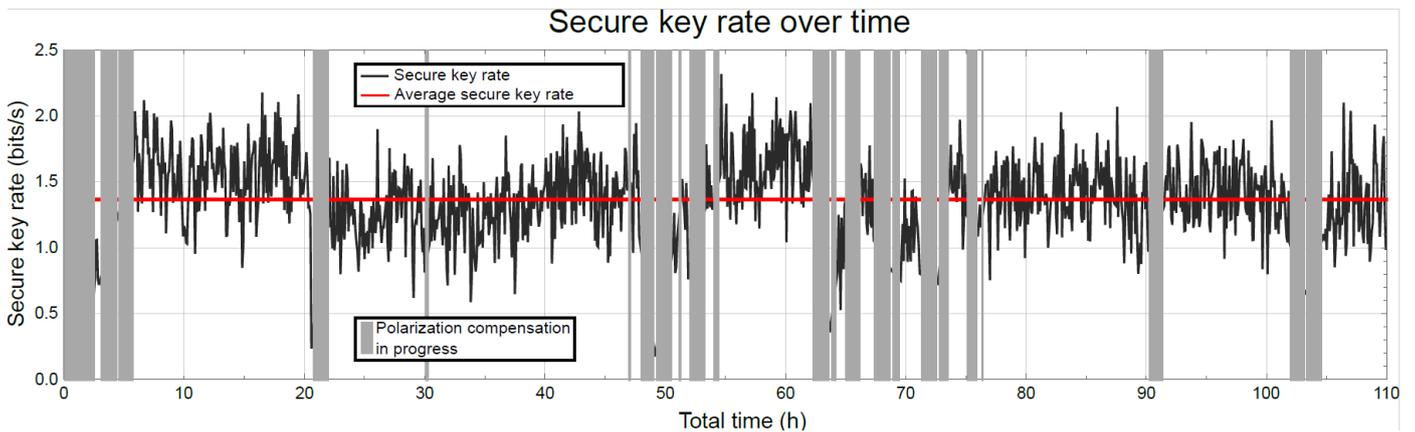

*Figure 3. Secure key rate over time. As long as the quantum bit error rate (QBER) stays below 11 %, a quantum secure key can be created in principle (see Fig. 2). Our polarization alignment procedure allows to keep the QBER in the key creation regime for altogether 82 out of 110 hours of total link operation. The red line gives the average secret key rate (1.4 bits/s) calculated from all coincidences over these 82 hours. Thus, the total secure key amounts to 403 kbit, calculated with the overall, i.e. average, QBER. In black we show the secure key rate based on data acquired over 300 second time windows. Its fluctuations in the order of ±0.3 bits/s originate from Poissonian photon statistics and polarization drifts.*

To analyse the performance in a QKD setting, one has to consider the number of coincident clicks as well as their QBER. For calculation of the final key rates, we follow the formula outlined in Ref.s [26, 33] (for details see the Methods section). Temporal filtering is of crucial importance for the final key size. Our optimal coincidence window $t_{CC}$ amounted to 114 ps. It maximizes the total key accumulated over 110 hours of link operation: 3.1 Mbit of raw key with a QBER of 7.0% yield 403 kbit of quantum secure key, equivalent to a rate of 1.4 bits/s during the active time of 82 hours (see Fig. 3). The obtainable maximum key rate will, due to finite-key effects, also depend on the actual running time of the link and might actually be lower if the link is operated for a short time only. To the best of our knowledge, the running time is only limited by the SNSPD maintenance cycle of about 10,000 hours.

**Discussion and conclusion**

We have shown an ultra-stable in-fibre polarization-based entanglement distribution scheme capable of creating quantum secure keys over a length of 248 km and a time span of 110 hours, overcoming a total 79 dB of loss along two nearly symmetric fibre links. To this end, we deploy a high-brightness, high-fidelity source of entangled photon pairs at telecommunication wavelengths together with high-end SNSPD systems. We operate the link at the current limit of the state-of-the-art, which mainly originates from the precision of timing synchronization. We manage to lower it to 114 ps by non-locally compensating for the dispersion of our 100 GHz WDM channels and by the use of SNSPDs. We find coincidence rates in the order of 7.7 $s^{-1}$ to be optimal to, on one hand, overcome loss and allow for live compensation of polarization drifts, and on the other hand to keep accidental coincidence rates sufficiently low.

Possible enhancements of our experiment could be realized by the use of several multiplexed wavelength channels, which could potentially increase the total key rates further [34]. Secondly, detection systems with lower detector jitter could allow for even stricter temporal filtering, which in turn enables higher pair production rates while keeping accidental coincidences low [27]. Thirdly, the PPC algorithm could be accelerated by automatically increasing the pump power during polarization alignment, which would yield better statistics for the QBER assessment. It might also allow for more stringent optimization, lowering the polarization-induced QBER below 1%. Fourthly, detailed trade-off calculations balancing the length of the polarization alignment with the quality of entanglement it establishes have to be carried out to determine optimal operation parameters. Fifthly, our entanglement distribution system could be integrated in wavelength-multiplexed quantum networks [16], e.g. by implementing additional short fibre links to several users in Vienna. Sixthly, our analysis has mainly focused on QKD. The performance regarding other implementations, e.g. quantum computation or blind computing, still have to be evaluated and might have far-reaching implications.

Summarizing, our work paves the way for all kinds of continuously operated applications of quantum entanglement distributed over long fibre-distances, most notably, but not limited to, quantum key distribution.

**METHODS**

**Source of entangled photon pairs**

The strong attenuation along both fibre links requires a high-brightness source of polarization-entangled photon pairs in order to achieve significant coincidence rates between the receivers. To this end, we deploy a Sagnac-type source based on spontaneous parametric down-conversion (SPDC) inside a bulk nonlinear ppLN crystal of type-0 phasematching pumped with a 775.06 nm continuous-wave Toptica laser. We choose a strong focusing parameter [31, 35] of $\xi = 1.99$ in order to arrive at pair production rates of $2.5 \times 10^6$ s$^{-1}$/nm/mW (before all losses). The spatially degenerate entangled photon pairs are separated from the pump via a dichroic mirror and a longpass filter and coupled into one single-mode fibre. From the source's 100 nm broad spectrum centred around 1550.12 nm, we select two 100 GHz wavelength division multiplexing (WDM) channels, using in-fibre add-drop multiplexers. The channel to SP (B) is cantered at 1550.92 nm (1549.32 nm). Photons in one channel are entangled with their partner in the other channel due to energy conservation in the SPDC process [22]. The source was operated at 422 mW pump power, producing $6.4 \times 10^8$ photon pairs per second. The source's intrinsic QBER due to erroneous polarization measurements was determined to be less than 0.4% in a laboratory environment.

**Single-photon polarization measurement**

Two fibre links connect the source of entangled photons, located in the basement of the University of Vienna's physics institute, to two measurement stations: A Türk Telekom repeater station in Getzersdorf, part of District St. Pölten in Lower Austria (SP), and Bob at the Research Center for Quantum Information on the campus of the Slovakian Academy of Sciences in Bratislava (B). The measurement apparatuses at

SP and B are identical in construction (see Fig. 1). They each consist of a bulk polarization measurement module (PMM), a 4-channel superconducting nanowire single-photon detector (SNSPD) and time-tagging electronics (TTM). Local measurements in the laboratory without long-distance link but including all losses in source, PMM and SNSPD, have shown heralding efficiencies of about 20% on average, equivalent to −7.0 dB. All additional loss in our experiment can be attributed to the fibre links and compensation stages.

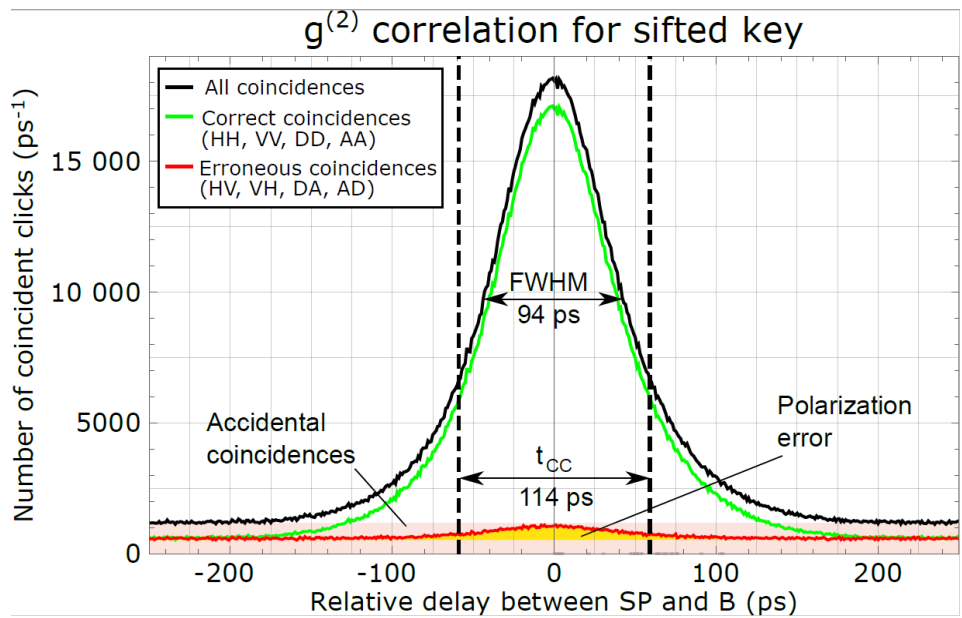

*Figure 4. $g^{(2)}$ intensity correlation of all coincidences used in key creation over the 82 hours of stable QBER. The dotted line shows the width of the coincidence window, $t_{CC}$ = 114 ps, yielding the highest total key of 403 kbit. The ratio of the areas below the red and below the black curve within $t_{CC}$ corresponds to the overall QBER of 7.0%. The yellow area depicts the number of erroneous measurements due to imperfect polarization alignment, amounting to about 2.6%. The remaining 4.4% of QBER can be attributed to accidental coincidences. We have subtracted the overall delay of approximately 44.8 µs from all relative delay values.*

In the PMM, the photons are coupled out of the long-distance fibre and impinge on a 50:50 beamsplitter randomly directing them to two mutually unbiased linear polarization basis measurements. The first basis is realized by a PBS transmitting (reflecting) the horizontal (vertical) polarization mode, which is then coupled into one single-mode fibre each. The other basis, measuring the diagonal/antidiagonal basis, works alike except for an HWP set to 22.5° before the PBS, effectively rotating the polarization modes by 45°. The four PMM single-mode fibres are connected to the SNSPD, which detects photons with a probability of ≈ 80 % according to the manufacturer Single Quantum. All detection events are recorded using a TTM by Swabian Instruments with 1 ps resolution. The combined jitter of SNSPD and TTM on both sides amounts to ≈ 38 ps full-width at half maximum (FWHM). In order to identify detector clicks at SP and B originating from the same photon pair, each TTM is disciplined to a GPS clock. The relative drift of these clocks, on average 13 ps/s, limits the maximum integration time over which detection events at both receivers can be acquired and compared. As can be seen in Fig. 4, the polarization measurement error along the link (2.6 %) is higher than in laboratory measurements (0.2 %) mainly due to our efforts to keep the PPC alignment time low, which did not allow us to set the entangled state perfectly for every alignment.

**Fibre link**

Measurements with an optical time domain reflectometer (OTDR) of the fibre to St. Pölten (Bratislava) yielded a fibre length L of 129.0 km (119.2 km) and losses of −31.9 dB (−32.6 dB). Additional to loss, there are two dispersion effects detrimental to QKD imposed by long-distance fibres: chromatic dispersion (CD) and polarization mode dispersion (PMD).

CD is proportional to L and to the signal's spectral width. It induces different travel times along the fibre for different parts of the light spectrum. This can effectively be seen as a decrease in temporal measurement precision, which smears out the correlation function between Alice and Bob, thus increasing the QBER and rendering live tracking impossible in our high-loss setting. In our experiment, we benefit from the fact that entanglement-based QKD allows for non-local dispersion compensation [28-30]. This means that the total CD effect of both fibre links can be reduced to zero by use of just one dispersion compensation module (DCM). It introduces additional attenuation of about −7 dB. OTDR measurements yielded a CD of 2073 ps/nm (617 ps/nm) to Alice (Bob). These values are so different because the fibre link to Bob partially conforms to the G.655 ITU standard, which allows less dispersion (6.0 ps/nm) than the more commonly used G.652 standard that was used for the link to Alice (16.8 ps/nm) [36]. Since about 24 km of fibre had not yet been connected to the link at the time the CD measurements were taken, we estimated the final overall CD to be 3.0 ns/nm, equivalent to 1.8 ns for our 100 GHz WDM spectra (with a FWHM of about 75 GHz ≈ 0.6 nm), and chose the DCM accordingly. The total FWHM of the correlation peaks amounts to 94 ps on average (see Fig. 4). We consider it a convolution of independent timing uncertainty effects: 38 ps originate from SNSPD and TTM jitter, as confirmed in the laboratory without link. The remaining 86 ps can be attributed either to the mean relative GPS clock drift accumulated over the post-processing integration time of 7s [37], or to residual uncompensated CD, or to both. In our realization, we had no means to differentiate between the two effects.

PMD causes different travel speeds for different polarization states inside the fibre. This is due to varying birefringence over the full stretch of the connection, which is in turn induced by random fibre imperfections. This effect scales with √L since the accumulated imperfections can not only add up, but also cancel each other [38]. If PMD induces a temporal delay between two orthogonal polarization states which is larger than an unpolarized photon's coherence time, it becomes polarized. In our case, this is equivalent to a polarization measurement and would therefore inhibit distribution of polarization entanglement [39]. OTDR measurements of the fibre to SP (B) have shown the PMD to be 0.63 ps (0.24 ps), which is substantially lower than our photons' estimated coherence time of ≈ 10 ps. Accordingly, we could not observe PMD-induced loss of polarization fidelity along our fibre link.

**Data acquisition**

Data was taken over the course of 109:55 hours, starting on December 14th, 2021 at 17:38 and ending on December 19th, 7:05. The average count rates of all four detectors combined in SP (B) amounted to 62,500 s$^{-1}$ (94, s$^{-1}$), where 1,200 s$^{-1}$ each originate from detector-intrinsic dark counts. Of these detector clicks, about 4.4 s$^{-1}$ were coincident in the same measurement bases ("sifted key") and can be used for key creation after error correction and privacy amplification. This number depends on the chosen coincidence window t$_{CC}$: For live operation, we chose t$_{CC}$ = 300 ps and an integration time of 9,600 ms in order to register as many coincidences as possible. Note that these are the parameters used for polarization alignment and not those used for key creation, since for live operation, loss of tracking has to be prevented at all cost in order to ensure polarization stabilization. Such stable live operation of the system however relies on automatic temporal tracking of the coincidence peak, which is moving in time due to the GPS clocks' relative temporal drift. If insufficient statistics, i.e. too few coincidences for the chosen integration time, cause the peak to be unrecognizable to the tracking algorithm, it can move out of the 1 ns monitoring window and be lost. This results in failure of the protocol. On the other hand, if the integration time is too long, the clock drift can already start to smear out the coincidence peak, and no additional precision can be gained by integrating further. We chose above parameters because they proved to work sufficiently well to not lose the tracking over the full measurement period, while still providing sufficient contrast for alignment. Thoroughly calculating the discussed trade-offs and optimizing the algorithm with regard to speed and effectiveness will be the subject of future studies. For key creation, which is done in post-processing, t$_{CC}$ = 114 ps and an integration time of 7 seconds – resulting in a sifted key rate of 3.8 s$^{-1}$ and an average QBER of 7.0 % – was shown to yield the largest key. For a detailed analysis of the choice of the optimal coincidence window, we refer the reader to the section "Key rate calculation" and Ref. [27].

**Active polarization stabilization**

Polarization drift along the deployed fibres due to stress, vibrations and temperature changes constitutes a challenge we overcome with the use of non-local polarization control. There have been approaches to automatize polarization drift compensation in both entanglement-based [40] and prepare-and-send [41, 42] implementations, which however operated in regimes of substantially less loss and on shorter timescales. Our scheme was implemented in the B fibre, right after the source, via one piezoelectric-crystal-based polarization controller module (PPC). We align with respect to the QBER directly. No additional equipment such as a time- or wavelength-multiplexed reference laser have to be used in this scheme, which greatly reduces the engineering overhead of the experiment. Polarization drifts in both fibre links could be compensated with just one PPC due to the non-local nature of our entangled state. The algorithm in use optimized the visibility of the entangled state by iterative scanning of the voltages applied to the PPC's four fibre-squeezing piezo-electric crystals. Since the coincidence rates used for live-tracking were only in the order of 5.3 $s^{-1}$, the most time-consuming part of polarization optimization is accumulating enough statistics to determine the current visibility value with sufficient precision, which we chose to be ±0.2 QBER percent points, assuming Poissonian statistics. As can be seen in Fig. 2, the initial alignment takes much longer than the later corrections (more than 2:30 hours). Subsequent alignments on the other hand can take as few as 8 minutes, and 57 minutes on average. We assume this is because the phase of the entangled state is completely random in the beginning. Polarization drifts, however, do not suddenly randomize the entangled state's relative phase but only cause gradual changes which can be compensated faster.

We managed to compensate for only 25.4% of the total time, which is important since the coincidence data used for polarization alignment can naturally not be used for key creation. This also means that continuous monitoring of the QBER value, which we did for illustrative purposes in this publication, is not possible. Therefore, one can check the QBER values at certain points in time only, e.g. every hour. This would mean that the actual duty cycle of the system is reduced by another 3.1 percent points, if one assumes the determination of the QBER to take about 2:30 minutes. Additionally, if the QBER is found to be above the PPC limit, it might be beneficial to discard all data collected since the last QBER measurement in order to not dilute the overall raw key. This however depends on an estimate of how fast the QBER actually changes, and on how great the violation of the PPC limit was. Such detailed trade-offs will be the subject of future studies.

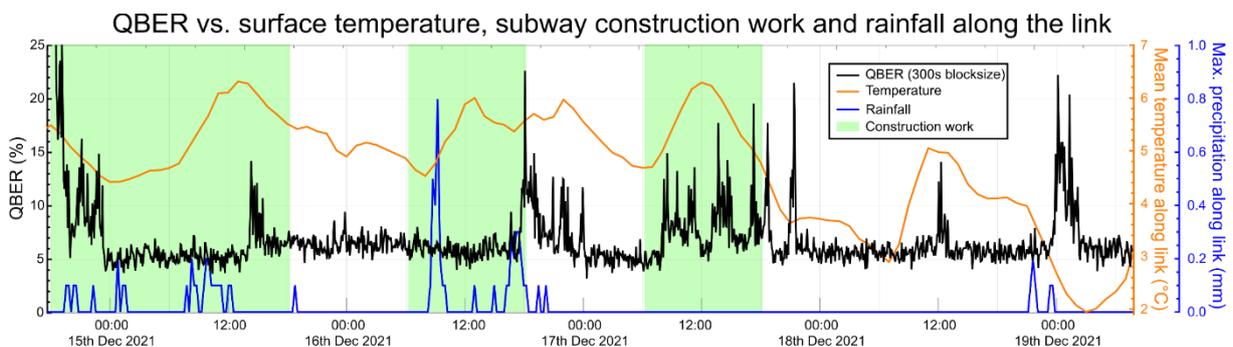

*Figure 5. Depiction of QBER over time along with weather and construction site data. There was little variation in mean air temperature along the link and only short stretches of light rain. The construction site, whose operation times are depicted in green, was located close to the deployed fibre connecting the source with the overland fibre links. It involved subterranean drilling for the construction of a new subway line. We can find no convincing evidence for influences from weather and construction work on the polarization stability along the link. However, we cannot exclude destabilizing effects of possible additional construction sites along our 248 km link, since we had no data about them available. Also, harsher weather conditions might still result in polarization drift.*

We also performed a preliminary investigation of possible environmental effects on the polarization stability (see Fig. 5). We compared QBER drifts with weather data by the Austrian Central Institution for Meteorology and Geodynamics (ZAMG) as well as with the working times at drilling sites for the future Viennese subway line U5, which we were supplied with by the Viennese public transport organization Wiener Linien. We find no convincing evidence for a correlation of any of above effects and polarization drift.

**Key rate calculation**

The key rate depends on several experimental factors. Parameters like pump power, link loss, chromatic dispersion, fidelity and stability of the quantum state, coupling efficiency of the source, and the coincidence window have been carefully analysed resp. chosen, following the methods outlined in Ref. [27]. For our calculation, we follow the key rate formula [33]

$$R^s = R^{sift}[1 - 2H_2(E)] \qquad (2)$$

where $R^s$ is the final secure key rate, $R^{sift}$ is the sifted key rate, i.e. the rate of coincidences measured in compatible bases, and $H_2(x)$ is the binary Shannon Entropy. Since both $R^{sift}$ and E depend on $t_{CC}$, one has to choose a coincidence window that neither excludes too many entangled pairs nor includes too many accidental counts (see Fig. 4).

The trade-off behind this calculation can be understood as follows: On one hand, if $t_{CC}$ is too big, we unnecessarily include uncorrelated accidental counts in our valid coincidences, thus increasing the QBER and losing key to error correction and privacy amplification. On the other hand, if $t_{CC}$ is chosen too small, too many valid coincidences get lost, and the size of our raw key decreases. This trade-off with regard to $t_{CC}$ further depends on the block size and the assumed error correction efficiency f. The latter we set to 1 in Eq. (2), since we assume an arbitrarily long key in our ultra-stable, actively compensated QKD scheme. If one however wants to divide the raw key into smaller blocks, f will increase [43], which would in turn also lead to a different optimal value of $t_{CC}$.

Such smaller block sizes might be desirable due to the fact that in a real-life implementation, the QBER unavoidably fluctuates in time. In this case, it is beneficial to split a block of QBER E into n sub-blocks with $E_1,...,E_n$ where $E = (\sum_{i=1}^{n} E_i)/n$ and calculate $R^s$ as the sum of all $R^s_i (E_i)$, since $H_2(x)$ is a concave function. On the other hand, smaller block sizes lead to less precise QBER estimates, effectively increasing the QBER, because one has to assume the worst E possible to guarantee a quantum secure key. Such trade-offs however are outside of the scope of this paper; we just present one final key size of 403 kbit for a single block containing the complete raw key, i.e. not exploiting above considerations for shorter blocks.

...

## Contributions

SPN, MB and RU designed the experiment. SPN and RU acquired the link infrastructure. SPN built the source of entangled photon pairs, the detection modules and the passive compensation stages. SPN and AB conducted the experiment. LB established the timing synchronization and wrote and operated the coincidence, data collection and data analysis software. AB wrote and operated the PPC algorithm. LB and AB built the PPC stage. SPN and MB wrote the paper.

## Data availability

The time-tag files collected in the course of the experiment are in the order of Terabyte and available from the corresponding authors upon reasonable request.

## Code availability

The code used for the polarization alignment procedure is available from the corresponding authors upon reasonable request.

## Competing interests

The authors declare no competing interests.

## Acknowledgements

We acknowledge European Union's Horizon 2020 programme grant agreement No.857156 (OpenQKD) and the Austrian Academy of Sciences in cooperation with the FhG ICON-Program "Integrated Photonic Solutions for Quantum Technologies (InteQuant)". We thank Peter Rapčan, Djeylan Aktas, Mário Ziman, and Vladimír Bužek of Bratislava for their help. We thank Ewald Martinelli, Harald Zeitlhofer and Giuseppe Antonelli of Türk Telekom for their help and discussions regarding the deployed fibre link. We thank Thomas Scheidl and Sören Wengerowsky for fruitful discussions and calculations, and Matej Pivoluska for advice regarding key rate estimates. We thank Ulrich Galander for prototyping parts of the experiment. We thank the Austrian Central Institution for Meteorology and Geodynamics (ZAMG) for the weather data and the Viennese public transport institution Wiener Linien for disclosing their construction site schedules.